\documentstyle[12pt,epsf]{article}

\begin{document}

\begin{titlepage}

\begin{flushright}
CERN-TH/97-3\\
hep-ph/9701375
\end{flushright}

\vspace{1cm}
\begin{center}
\Large\bf
Two-Loop Anomalous Dimension of the\\
Chromo-Magnetic Moment\\
of a Heavy Quark
\end{center}

\vspace{1.5cm}
\begin{center}
G. Amor\'os$^*$, M. Beneke and M. Neubert\\[0.1cm]
{\sl Theory Division, CERN, CH-1211 Geneva 23, Switzerland}
\end{center}

\vspace{1cm}
\begin{abstract}
\vspace{0.2cm}\noindent
We find that the anomalous dimension (in the $\overline{\mbox{MS}}$
scheme) of the chromo-magnetic operator $g_s\bar h_v\sigma_{\mu\nu}
G^{\mu\nu} h_v$ in the heavy-quark effective theory is
\[
   \phantom{ \Bigg[ }
   \gamma_{\rm mag} = \frac{C_A\alpha_s}{2\pi}\left[ 1
   + \left( \frac{17}{18}\,C_A - \frac{13}{18}\,T_F\,n_f \right)
   \frac{\alpha_s}{\pi} + O(\alpha_s^2) \right] \,,
\]
generalizing the one-loop expression known previously. To derive the
two-loop result, we use the reparametrization invariance and the
virial theorem. After performing infrared subtractions, all two-loop
integrals are of propagator type and are evaluated by a recurrence
relation for tensor integrals.

\end{abstract}

\vspace{1cm}
\centerline{(Submitted to Physics Letters B)}

\vspace{3cm}
\noindent
CERN-TH/97-3\\
January 1997

\vspace{1.5cm}
\centerline{$^*$\small
On leave from: Departament de F\'\i sica Te\`orica, Universitat de
Val\`encia, Spain}

\end{titlepage}

\section{Introduction}

The heavy-quark effective theory (HQET) is a convenient tool to
describe the physics of hadrons containing a heavy quark
\cite{review}. It provides a systematic expansion around the limit
$m_Q\to\infty$, in which new symmetries of the strong interactions
arise, relating the long-distance properties of many observables to a
small number of hadronic matrix elements. The effective Lagrangian of
the HQET is \cite{Geor}--\cite{FGL}
\begin{equation}
   {\cal L}_{\rm eff} = \bar h_v\,i v\!\cdot\!D\,h_v
   + \frac{C_{\rm kin}}{2 m_Q}\,\bar h_v(i D)^2 h_v
   + \frac{C_{\rm mag}\,g_s}{4 m_Q}\,
   \bar h_v\sigma_{\mu\nu} G^{\mu\nu} h_v + O(1/m_Q^2) \,,
\label{Leff}
\end{equation}
where $g_s G^{\mu\nu}=i[D^\mu,D^\nu]$ is the gluon field-strength
tensor, and $h_v$ the velocity-dependent field describing a heavy
quark inside a hadron moving with velocity $v$. This field is
subject to the constraint $\rlap/v\,h_v=h_v$. The leading term in the
effective Lagrangian, which gives rise to the Feynman rules of the
HQET, is invariant under a global $SU(2n_h)$ spin--flavour symmetry
group, where $n_h$ is the number of heavy-quark flavours. This
symmetry is broken by the higher-dimensional operators arising at
order $1/m_Q$. The first operator corresponds to the kinetic energy
of the heavy quark inside the hadron, and the second operator
describes the magnetic interaction of the heavy-quark spin with the
gluon field. The coefficients $C_{\rm kin}$ and $C_{\rm mag}$ result
from short-distance effects and, in general, depend on the scale at
which the operators are renormalized.

Hadronic matrix elements of the higher-dimensional operators in
(\ref{Leff}) play a significant role in many applications of the
HQET. It is therefore important to study the properties of these
operators, in particular under renormalization. The reparametrization
invariance of the HQET (an invariance under infinitesimal changes of
the velocity) implies that, in regularization schemes with a
dimensionless regulator, the kinetic operator is not renormalized,
i.e.\ $C_{\rm kin}=1$ to all orders in perturbation theory
\cite{LuMa}. (In regularization schemes with a dimensionful
regulator, the kinetic operator mixes with lower-dimensional
operators \cite{Chris,inv2}.) The renormalization of the
chromo-magnetic operator at the one-loop order is known for some time
\cite{EiHi,FGL}. It is governed by the anomalous dimension, defined
as
\begin{equation}
   \gamma_{\rm mag} = Z_{\rm mag}\,
   \frac{\mbox{d}Z_{\rm mag}^{-1}}{\mbox{d}\ln\mu}
   = 2\alpha_s\,\frac{\partial}{\partial\alpha_s}\,
   Z_{\rm mag}^{(1)} \,,
\label{gammag}
\end{equation}
where $Z_{\rm mag}$ is the renormalization constant that relates the
bare operator with the renormalized one, $O_{\rm mag} = Z_{\rm
mag}\,O_{\rm mag}^{\rm bare}$. The last relation is valid in
dimensional regularization with a MS-like subtraction scheme, such as
$\overline{\mbox{MS}}$ \cite{BAR}, and follows from the requirement
that the anomalous dimension be finite as $\epsilon=(4-d)/2\to 0$
\cite{Flor}. $Z_{\rm mag}^{(1)}$ denotes the coefficient of the
$1/\epsilon$ pole in $Z_{\rm mag}$. At the one-loop order,
$\gamma_{\rm mag}=C_A\alpha_s/2\pi$ \cite{EiHi,FGL}. In this paper,
we calculate the anomalous dimension to two-loop order. This
completes the renormalization of the HQET Lagrangian at order
$1/m_Q$, and to next-to-leading order in $\alpha_s$.

A direct calculation of the renormalization factor $Z_{\rm mag}$ at
order $\alpha_s^2$ would require the evaluation of a large number of
two-loop diagrams. The equivalent calculation for the chromo-magnetic
operator of a light quark has been performed in Ref.~\cite{MiMu}. For
heavy quarks, we follow a different strategy, which exploits the
symmetries of the HQET. Instead of the chromo-magnetic operator
itself, we study the renormalization of the two operators
\begin{equation}
   O_1^{\mu\nu} = g_s\bar h_v\Gamma\,iG^{\mu\nu} h_v \,, \qquad
   O_2^{\mu\nu} = v^\mu v_\rho O_1^{\rho\nu}
    - v^\nu v_\rho O_1^{\rho\mu} \,.
\label{basis}
\end{equation}
For the special choice $\Gamma=-i\sigma_{\mu\nu}$, the first operator
reduces to the chromo-magnetic operator, while the second one
vanishes. Because the Feynman rules of the HQET do not involve
$\gamma$ matrices, the Dirac structure of $\Gamma$ is not altered by
radiative corrections, and it is of advantage to treat it as a
general matrix. Then the operators $O_1^{\mu\nu}$ and $O_2^{\mu\nu}$
form a basis of physical (class-1) operators, i.e.\ operators that do
not vanish by the equations of motion. The mixing of these operators
under renormalization is such that \cite{Amor}
\begin{equation}
   O_1^{\mu\nu} = Z_1\,O_{1,{\rm bare}}^{\mu\nu}
    + Z_2\,O_{2,{\rm bare}}^{\mu\nu} \,, \qquad
   O_2^{\mu\nu} = (Z_1 + Z_2)\,O_{2,{\rm bare}}^{\mu\nu} \,.
\label{mix}
\end{equation}
In the rest frame of the heavy quark, the operator $O_2^{\mu\nu}$
contains only chromo-electric field components. The virial theorem of
the HQET \cite{virial} relates this operator to the kinetic operator
in (\ref{Leff}). Since the kinetic operator is not renormalized, we
have the exact relation $Z_1+Z_2=1$, i.e.\ \cite{Amor}
\begin{equation}
    Z_{\rm mag} = Z_1 = 1 - Z_2 \,.
\label{nice}
\end{equation}
Below we calculate $Z_2$ to order $\alpha_s^2$, and then use
(\ref{nice}) and (\ref{gammag}) to obtain the two-loop anomalous
dimension of the chromo-magnetic operator.

\section{Two-loop Calculation}
\label{sec:2}

The particular structure of the operator $O_2^{\mu\nu}$ greatly
facilitates the calculation. To obtain the factor $Z_2$ in
(\ref{mix}) at order $\alpha_s^2$, we calculate the insertion of
$O_1^{\mu\nu}$ into the amputated Green function with two heavy
quarks and a (background-field) gluon to two-loop order, as well as
the one-loop diagrams with counterterm insertions needed to subtract
the subdivergences of the two-loop diagrams. However, we only need to
evaluate those diagrams giving a contribution proportional to the
tree-level matrix element of $O_2^{\mu\nu}$:
\begin{equation}
   \langle O_2^{\mu\nu}\rangle = g_s t_a \Gamma \left[
   v^\alpha (q^\mu v^\nu - q^\nu v^\mu)
   + v\cdot q\,(v^\mu g^{\alpha\nu} - v^\nu g^{\alpha\mu}) \right]
\,,
\label{O2str}
\end{equation}
where $v$ is the heavy-quark velocity, $t_a$ a colour matrix,
$\alpha$ and $a$ the Lorentz and colour indices of the gluon, and $q$
the incoming gluon momentum. It suffices to trace in our calculation
the terms proportional to $v^\alpha v^\nu q^\mu$, as no other
operators yield this Lorentz structure. Then, taking into account the
Feynman rules of the HQET, it is straightforward to derive the
following selection rules:
\begin{enumerate}
\item
If the external gluon is attached to the operator, the diagram is
proportional to $g^{\mu\alpha}$ or $g^{\nu\alpha}$ and does not
contribute to the structure $v^\alpha v^\nu q^\mu$.
\item
If the external gluon is attached to a heavy-quark line, the diagram
only depends on $v\cdot q$ and does not contribute to the desired
structure. On the other hand, the external gluon must be connected to
a heavy-quark line through other gluons (or through a light-fermion
or ghost loop), otherwise one cannot get $v^\alpha$.
\item
The external gluon must be connected to the operator through other
gluons (or through a light-fermion or ghost loop). If it is not, one
cannot get $q^\mu$.
\end{enumerate}

\begin{figure}
\epsfxsize=6cm
\centerline{\epsffile{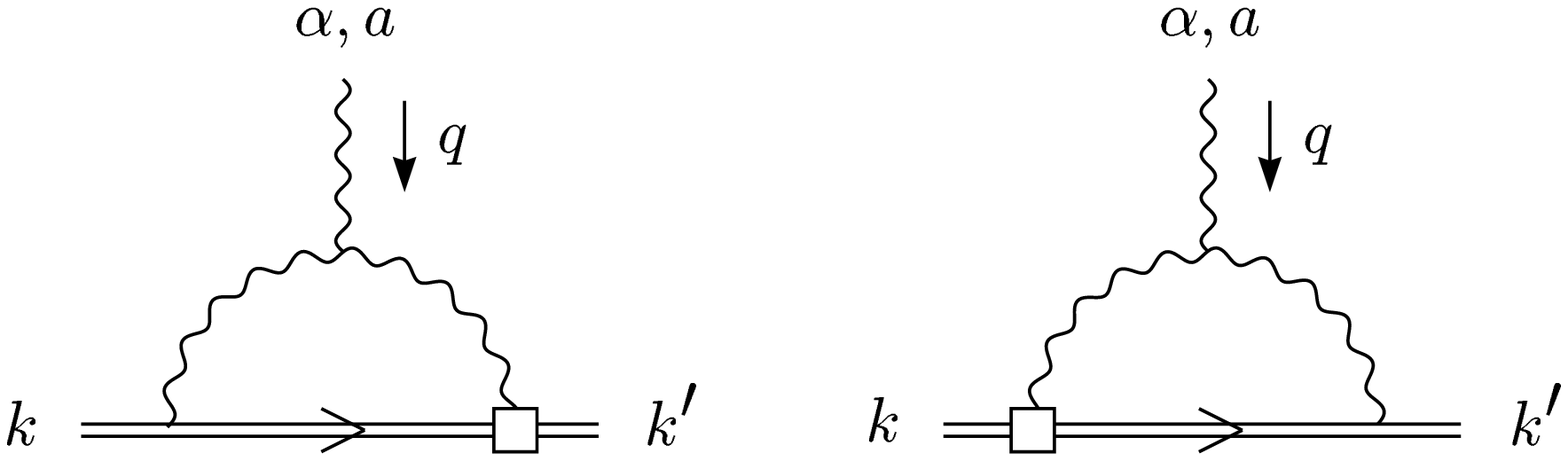}}
\vspace{-0.5cm}
\centerline{\parbox{14cm}{\caption{\label{fig:1loop}
One-loop diagrams contributing to the calculation of the
renorma\-li\-za\-tion factor $Z_2$. The operator $O_1^{\mu\nu}$ is
represented by a square.}}}
\end{figure}

\begin{figure}
\epsfxsize=15cm
\centerline{\epsffile{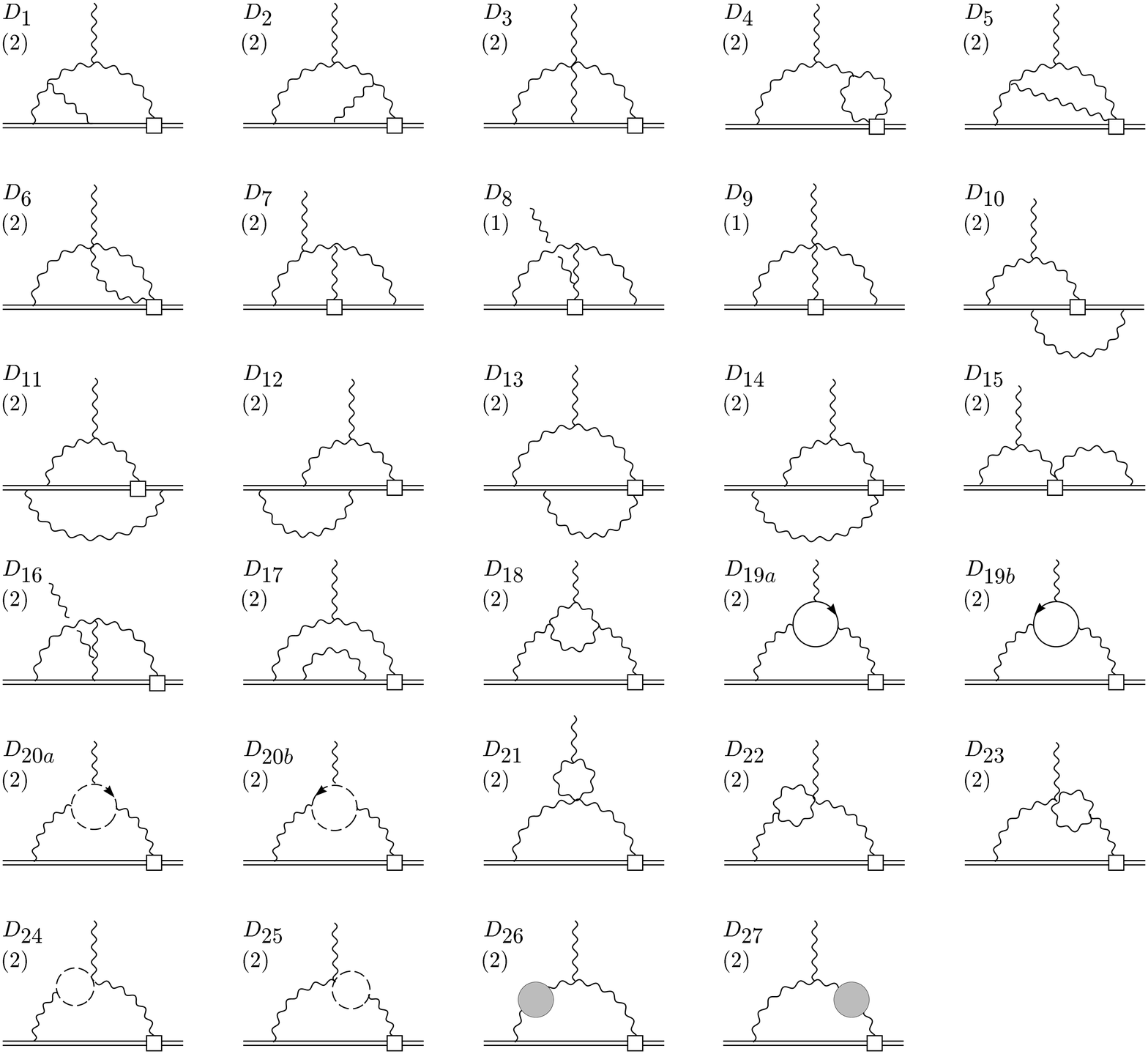}}
\vspace{-0.4cm}
\centerline{\parbox{15cm}{\caption{\label{fig:2loop}
Two-loop diagrams contributing to the calculation of the
renormalization factor $Z_2$. The notation $(2)$ indicates that a
mirror copy of the diagram is included implicitly. The shaded circles
represent one-loop self-energy insertions.}}}
\end{figure}

Accounting for these rules, only the two diagrams shown in
Fig.~\ref{fig:1loop} need to be calculated at the one-loop order.
They give identical contributions. At the two-loop order, all
diagrams that yield terms proportional to $v^\alpha v^\nu q^\mu$ can
be generated by first drawing the two-loop diagrams that contribute
to the heavy-quark two-point function with an insertion of
$O_1^{\mu\nu}$, and then attaching an external gluon in all possible
ways to one of the gluon, ghost or light-quark lines or vertices.
This gives rise to the graphs shown in Fig.~\ref{fig:2loop}.

In evaluating the two-loop diagrams, we only need to keep terms
linear in the gluon momentum $q$. Because the pole parts are
polynomial in the external momenta, we can first take one derivative
with respect to $q$ and then set $q=0$ and $k=k'$, so that all
integrals are of propagator type and depend on the single variable
$\omega=v\cdot k=v\cdot k'$, where $k$ and $k'$ are the external
momenta of the heavy quarks. However, this straightforward
application of the method of infrared (IR) rearrangement \cite{VL1}
fails for some of the diagrams ($D_2$, $D_5$, $D_7$, $D_{13}$,
$D_{15}$, $D_{18}$ and $D_{21}$), for which setting $q=0$ after
differentiation leads to IR divergences. In these cases, we apply a
simple variant of the so-called $R^*$ operation \cite{Che}, which
compensates these IR poles by a recursive construction of
counterterms for the IR divergent subgraphs. In our case, after
accounting for UV and IR counterterms in all subgraphs, the overall
counterterm is purely UV and gives rise to the desired contribution
to the anomalous dimension.

\begin{figure}
\epsfxsize=14cm
\centerline{\epsffile{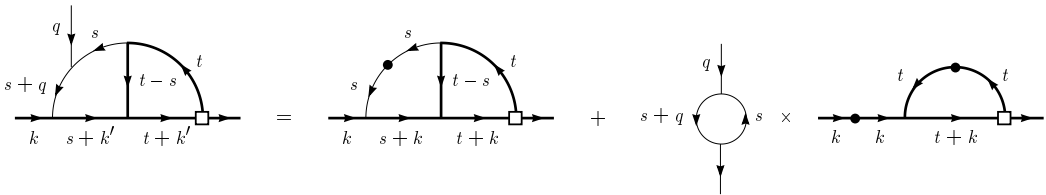}}
\centerline{\parbox{15cm}{\caption{\label{fig:rstar}
Schematic representation of the $R^*$ operation. Thick (thin) lines
show propagators with a large (small) momentum flow. The black dots
represent the original vertices.}}}
\end{figure}

To illustrate the procedure, consider a typical integral, which
arises
(as coefficient of $q^\mu$) in the calculation of $D_2$:
\begin{equation}
   I = \int\mbox{d}^ds\,\mbox{d}^dt\,\frac{t^\beta t^\gamma}
   {(v\cdot s+\omega)(v\cdot t+\omega)(-s^2)[-(s+q)^2](-t^2)
    [-(t-s)^2]} \,.
\label{Iint}
\end{equation}
For $q=0$, this integral has an IR divergence when $s\to 0$. To
construct the IR counter\-term that compensates this singularity, we
subtract and add the integral
\begin{equation}
   I_{\rm IR} = \int\mbox{d}^ds\,\frac{1}{(-s^2)[-(s+q)^2]}\,
   \int\mbox{d}^dt\,\frac{t^\beta t^\gamma}
   {\omega\,(v\cdot t+\omega)(-t^2)^2} \,,
\end{equation}
which is obtained by setting $s=0$ whenever possible (i.e.\ by
neglecting $s$ compared with $t$, and $v\cdot s$ compared with
$\omega$). The two integrals $I$ and $I_{\rm IR}$ have the same
behaviour for $s\to 0$; their difference is IR finite and can be
evaluated for $q=0$. In this case $I_{\rm IR}$ becomes a tadpole
integral and vanishes, so that $(I - I_{\rm IR})_{q=0}$ coincides
with the original integral with $q$ set to zero. This integral can be
calculated using the general algorithm described below. On
dimensional grounds, it is proportional to $(-2\omega)^{-4\epsilon}$.
The contribution $I_{\rm IR}$, which is necessary to subtract the IR
subdivergence of the original integral $I$, factorizes into an IR
counterterm (the $s$ integral) and the original diagram with the
lines of the IR sensitive subgraph removed. This deletion of lines
follows from the locality of IR divergences in momentum space.
Because of this factorization, IR subtraction terms such as $I_{\rm
IR}$ are proportional to $(-2\omega)^{-2\epsilon}
(-q^2)^{-\epsilon}$. The construction just described is schematically
shown in Fig.~\ref{fig:rstar}. It is necessary that exactly the same
treatment is followed for the one-loop UV counterterms associated
with a given diagram. Then the $1/\epsilon$ poles proportional to
$\ln(-2\omega/\mu)$ and $\ln(-q^2/\mu^2)$ are cancelled for each
diagram, and the final overall counterterm is local.

All integrals can be treated in the above manner. The remaining
two-loop tensor integrals are of the general form
\begin{eqnarray}
   &&\int\mbox{d}^ds\,\mbox{d}^dt\,
   \left( \frac{\omega}{v\cdot s+\omega} \right)^{\alpha_1}
   \left( \frac{\omega}{v\cdot t+\omega} \right)^{\alpha_2}
   \frac{s_{\mu_1}\dots s_{\mu_n} t^{\nu_1}\dots t^{\nu_m}}
    {\big(-s^2\big)^{\alpha_3} \big(-t^2\big)^{\alpha_4}
     \big[-(s-t)^2\big]^{\alpha_5}}
    \nonumber\\
   &\equiv& -\pi^d (-2\omega)^{2(d-\alpha_3-\alpha_4-\alpha_5)+n+m}\,
    I_{\mu_1\dots\mu_n}^{\nu_1\dots\nu_m}(\{\alpha_i\}) \,.
\end{eqnarray}
The diagrams $D_{10}$ and $D_{12}$ can be brought into this form by
multiplication with
\begin{equation}
   1 = \frac{1}{\omega} \Big\{ (v\cdot s+\omega)
   + (v\cdot t+\omega) - [v\cdot(s+t)+\omega] \Big\} \,.
\end{equation}
A standard method is to reduce the tensor integrals to scalar
two-loop integrals, which can be calculated in an algorithmic way
\cite{BrGr}. However, since in our case we need integrals with up to
five tensor indices, the reduction to scalar integrals would be
rather involved. Therefore, we compute the tensor integrals directly.
Using the method of integration by parts \cite{Chet}, we obtain the
recurrence relation
\begin{eqnarray}
   &&\left[ (d-\alpha_1-\alpha_3-2\alpha_5+n)
    + \alpha_3\,{\bf 3^+} ({\bf 4^-} - {\bf 5^-})
    + \alpha_1\,{\bf 1^+ 2^-} \right]\,
    I_{\mu_1\dots\mu_n}^{\nu_1\dots\nu_m}(\{\alpha_i\}) \nonumber\\
   &&\quad = \sum_{j=1}^n\,
    I_{\mu_1\dots[\mu_j]\dots\mu_n}^{\mu_j\nu_1\dots\nu_m}
    (\{\alpha_i\}) \,,
\end{eqnarray}
which allows us to express any two-loop integral in terms of
degenerate integrals, which have $\alpha_2=0$, $\alpha_4=0$ or
$\alpha_5=0$. Here ${\bf 1^+}$ is an operator raising the index
$\alpha_1$ by one unit etc., and $[\mu_j]$ means that this index is
omitted. The degenerate integrals can be
expressed in terms of products of
one-loop tensor integrals, which are given by
\begin{eqnarray}
   \int\mbox{d}^ds\,
   \left( \frac{\omega}{v\cdot s+\omega} \right)^\beta
   \frac{s^{\mu_1}\dots s^{\mu_n}}{\big(-s^2\big)^\alpha}
   &=& i\pi^{d/2} I_n(\alpha,\beta)\,(-2\omega)^{d-2\alpha+n}\,
    K^{\mu_1\dots\mu_n}(v;\alpha) \,, \nonumber\\
   \int\mbox{d}^ds\,
   \frac{s^{\mu_1}\dots s^{\mu_n}}{\big(-s^2\big)^\alpha
    \big[-(s-q)^2\big]^\beta}
   &=& i\pi^{d/2} G_n(\alpha,\beta)\,(-q^2)^{d/2-\alpha-\beta}\,
    J^{\mu_1\dots\mu_n}(q;\alpha,\beta) \,,
\end{eqnarray}
with
\begin{eqnarray}
   I_n(\alpha,\beta) &=& \frac{\Gamma(d/2-\alpha+n)\,
    \Gamma(2\alpha+\beta-d-n)}{\Gamma(\alpha)\,\Gamma(\beta)} \,,
    \nonumber\\
   G_n(\alpha,\beta) &=& \frac{\Gamma(d/2-\alpha+n)\,
    \Gamma(d/2-\beta)\,\Gamma(\alpha+\beta-d/2)}
    {\Gamma(\alpha)\,\Gamma(\beta)\,\Gamma(d-\alpha-\beta+n)} \,.
\label{IGfun}
\end{eqnarray}
The tensor structures are
\begin{eqnarray}
   K^{\mu_1\dots\mu_n}(v;\alpha) &=& \sum_{j=0}^{[n/2]}\,
    (-1)^{n-j}\,C(n,j;\alpha)\,
    \sum_{\nu_i=\sigma(\mu_i)}\mbox{}\!\!\!\!\!'\,
    g^{\nu_1\nu_2}\dots g^{\nu_{2j-1}\nu_{2j}}\,
    v^{\nu_{2j+1}}\dots v^{\nu_n} \,, \nonumber\\
   J^{\mu_1\dots\mu_n}(q;\alpha,\beta) &=& \sum_{j=0}^{[n/2]}\,
    (-q^2)^j\,D(n,j;\alpha,\beta)\,
    \sum_{\nu_i=\sigma(\mu_i)}\mbox{}\!\!\!\!\!'\,
    g^{\nu_1\nu_2}\dots g^{\nu_{2j-1}\nu_{2j}}\,
    q^{\nu_{2j+1}}\dots q^{\nu_n} \,,
\label{tensors}
\end{eqnarray}
where $[n/2]$ is the largest integer less than or equal to $n/2$, and
\begin{eqnarray}
   C(n,j;\alpha) &=& \prod_{k=1}^j\,\frac{1}{d+2(n-k-\alpha)} \,,
    \nonumber\\
   D(n,j;\alpha,\beta) &=& \prod_{k=1}^j\,
    \frac{d+2(k-1-\beta)}{[d+2(n-k-\alpha)][d+2(k-\alpha-\beta)]} \,.
\end{eqnarray}
The primed sums in (\ref{tensors}) mean a summation over all
$\Big( \begin{array}{c} n \\ 2j \end{array} \Big)(2j-1)!!$
permutations that lead to a different assignment of indices.

\begin{table}
\caption{\label{tab:2loop}
Pole parts proportional to $g_s t_a\Gamma\,v^\alpha v^\nu q^\mu$ of
the two-loop diagrams in units of $(\alpha_s/4\pi)^2$. The diagrams
$D_6$, $D_8$, $D_{14}$, $D_{16}$, $D_{24}$ and $D_{25}$ vanish or do
not yield the desired Lorentz structure.}
\vspace{0.4cm}
\begin{center}
\begin{tabular}{|c|c|c|c|}\hline
\rule[-0.35cm]{0cm}{0.95cm}
 & colour factor
 & $\times\left( \frac{-2\omega}{\mu} \right)^{-4\epsilon}$
 & $\times\left( \frac{-2\omega}{\mu} \right)^{-2\epsilon}
   \left( \frac{-q^2}{\mu^2} \right)^{-\epsilon}$ \\
\hline
$D_1$ & $C_A^2$ & $\frac{1}{16\epsilon}$ & \rule{0cm}{0.5cm}
 \\[0.15cm]
$D_2$ & $C_A^2$ & $-\frac{1}{8\epsilon^2} - \left( \frac{101}{48}
                  - \frac{\pi^2}{9} \right) \frac{1}{\epsilon}$
      & $\phantom{-}\frac{1}{\epsilon^2}
        + \frac{6}{\epsilon}$ \\[0.15cm]
$D_3$ & $C_A^2$ & $\left( \frac{1}{6} - \frac{\pi^2}{9} \right)
                  \frac{1}{\epsilon}$ & \\[0.15cm]
$D_4$ & $C_A^2$ & $-\frac{3}{4\epsilon^2} - \frac{9}{8\epsilon}$
      & \\[0.15cm]
$D_5$ & $C_A^2$ & $-\frac{1}{2\epsilon^2} - \frac{7}{8\epsilon}$
      & $\phantom{-}\frac{1}{\epsilon^2}
        + \frac{2}{\epsilon}$ \\[0.15cm]
$D_7$ & $C_A^2$ & $\frac{11}{8\epsilon^2} + \left( \frac{85}{24}
                  - \frac{\pi^2}{9} \right) \frac{1}{\epsilon}$
      & $-\frac{1}{\epsilon^2} - \frac{6}{\epsilon}$ \\[0.15cm]
$D_9$ & $C_A^2$ & $\left( \frac{1}{3} + \frac{\pi^2}{9} \right)
                  \frac{1}{\epsilon}$ & \\[0.15cm]
$D_{10}$ & $C_A (C_A-2 C_F)$ & $\frac{1}{2\epsilon^2}
           + \frac{1}{\epsilon}$ & \\[0.15cm]
$D_{11}$ & $C_A (C_A-2 C_F)$ & $\frac{1}{2\epsilon^2}
           - \frac{1}{\epsilon}$ & \\[0.15cm]
$D_{12}$ & $C_A (C_A-2 C_F)$ & $\frac{1}{2\epsilon^2}
           + \frac{1}{\epsilon}$ & \\[0.15cm]
$D_{13}$ & $C_A^2$ & $\frac{1}{\epsilon^2} + \frac{2}{\epsilon}$
         & $-\frac{2}{\epsilon^2} - \frac{8}{\epsilon}$ \\[0.15cm]
$D_{15}$ & $C_A^2$ & $-\frac{2}{\epsilon^2} - \frac{4}{\epsilon}$
         & $\phantom{-}\frac{2}{\epsilon^2}
           + \frac{8}{\epsilon}$ \\[0.15cm]
$D_{17}$ & $C_A C_F$ & $\frac{1}{\epsilon^2} + \frac{2}{\epsilon}$
         & \\[0.15cm]

$D_{18}$ & $C_A^2$ & $\frac{5}{24\epsilon^2} + \frac{49}{72\epsilon}$
         & $\phantom{-}\frac{1}{\epsilon^2}
           + \frac{5}{2\epsilon}$ \\[0.15cm]
$D_{19}$ & $C_A T_F\,n_f$ & $\frac{2}{3\epsilon^2}
                           - \frac{5}{9\epsilon}$ & \\[0.15cm]
$D_{20}$ & $C_A^2$ & $-\frac{1}{24\epsilon^2}
                     - \frac{5}{72\epsilon}$ & \\[0.15cm]
$D_{21}$ & $C_A^2$ & & $-\frac{3}{\epsilon^2}
                       - \frac{6}{\epsilon}$ \\[0.15cm]
$D_{22}$ & $C_A^2$ & $-\frac{9}{16\epsilon^2}
                     - \frac{9}{32\epsilon}$ & \\[0.15cm]
$D_{23}$ & $C_A^2$ & $\frac{9}{16\epsilon^2}
                     + \frac{9}{32\epsilon}$ & \\[0.15cm]
$D_{26}$ & $C_A^2$ & $\frac{5}{6\epsilon^2} + \frac{17}{36\epsilon}$
         & \\[0.15cm]
         & $C_A T_F\,n_f$ & $-\frac{2}{3\epsilon^2}
                            - \frac{1}{9\epsilon}$ & \\[0.15cm]
$D_{27}$ & $C_A^2$ & $\frac{5}{6\epsilon^2} + \frac{47}{36\epsilon}$
         & \\[0.15cm]
         & $C_A T_F\,n_f$ & $-\frac{2}{3\epsilon^2}
                            - \frac{7}{9\epsilon}$ & \\[0.15cm]
\hline
\rule{0cm}{0.5cm}
Sum & $C_A^2$ & $\frac{7}{3\epsilon^2} + \frac{25}{18\epsilon}$
    & $-\frac{1}{\epsilon^2} - \frac{3}{2\epsilon}$ \\[0.15cm]
    & $C_A C_F$ & $-\frac{2}{\epsilon^2}$ & \\[0.15cm]
    & $C_A T_F\,n_f$ & $-\frac{2}{3\epsilon^2}
                      - \frac{13}{9\epsilon}$ & \\[0.15cm]
\hline
\end{tabular}
\end{center}
\end{table}

Using this technique, we have calculated the pole parts of the
two-loop diagrams shown in Fig.~\ref{fig:2loop}. We adopt the
background-field formalism \cite{tHof}--\cite{Abbo} and work in the
`t~Hooft--Feynman gauge. The final expression for the sum of all
diagrams is gauge independent. The pole parts proportional to
$v^\alpha v^\nu q^\mu$ are given in Tab.~\ref{tab:2loop} for each
diagram. Here $C_A=N$, $C_F=(N^2-1)/(2N)$, $T_F=1/2$ are the colour
factors for an $SU(N)$ gauge group, and $n_f$ is the number of
light-quark flavours. The renormalization scale $\mu$ is introduced
by the replacement of the bare coupling constant with the
renormalized one through the relation $g_s^{\rm
bare}=\bar\mu^\epsilon Z_g\,g_s$, with $\bar\mu=\mu\, e^{\gamma_E/2}
(4\pi)^{-1/2}$ in the $\overline{\mbox{MS}}$ scheme.

\section{Ultraviolet Counterterms}

The two-loop diagrams in Fig.~\ref{fig:2loop} contain subdivergences,
which must be subtracted by UV counterterms. We first discuss the
one-loop counterterms for the heavy-quark and gluon propagators and
vertices. The corresponding terms in the Lagrangian are
\begin{eqnarray}
   {\cal L}_{c.t.} &=& \delta_h\,\bar h_v\,iv\cdot\partial\,h_v
    - \frac{\delta_A}{2}\,(\partial_\mu A_\nu^a
    - \partial_\nu A_\mu^a)\,\partial^\mu A^{\nu a}
    + \bar\mu^\epsilon g_s\,\delta_{hhA}\,\bar h_v\,v\cdot A\,h_v
    \nonumber\\
   &&\mbox{}- \bar\mu^\epsilon g_s\,\delta_{BAA}\,f^{abc}
    \left[ (\partial_\mu B_\nu^a) A^{\mu b} A^{\nu c}
    + (\partial_\mu A_\nu^a) B^{\mu b} A^{\nu c}
    + (\partial_\mu A_\nu^a) A^{\mu b} B^{\nu c} \right] \,,
\end{eqnarray}
where $A^\mu$ is the conventional gluon field, and $B^\mu$ the
background field. In the Feynman gauge, the counterterm coefficients
are
\begin{eqnarray}
   \delta_h &=& Z_h - 1 = 2 C_F\,\frac{\alpha_s}{4\pi\epsilon} \,,
    \nonumber\\
   \delta_A &=& Z_A - 1 = \left( \frac 53\,C_A - \frac 43\,T_F\,n_f
    \right) \frac{\alpha_s}{4\pi\epsilon} \,, \nonumber\\
   \delta_{hhA} &=& Z_g Z_h Z_A^{1/2} - 1
    = (-C_A+2 C_F)\,\frac{\alpha_s}{4\pi\epsilon} \,, \nonumber\\
   \delta_{BAA} &=& Z_g Z_B^{1/2} Z_A - 1 = \delta_A \,,
\end{eqnarray}
where $Z_A$ is the wave-function renormalization factor for the gluon
field, and $Z_B^{1/2}=Z_g^{-1}$ holds for the background field
\cite{Abbo}. Note that the Feynman rule for the counterterm for the
three-gluon vertex with a background field is the same as that for
the usual three-gluon vertex, but with a different renormalization
factor.

In addition, the two-loop diagrams in Fig.~\ref{fig:2loop} contain
subdivergences that are removed by operator counterterms. To find
these counterterms, we calculate at the one-loop order all insertions
of the operator $O_1^{\mu\nu}$ into the amputated Green functions
with a non-negative degree of divergence. They have the field content
$\bar h_v h_v$, $\bar h_v h_v A$, $\bar h_v h_v B$, $\bar h_v h_v
AA$, $\bar h_v h_v AB$, $\bar h_v h_v BB$ and $\bar h_v
h_v\eta\bar\eta$, where $\eta$ is a ghost field. It turns out that
there are no UV divergences in the Green function containing ghost
fields, and that the insertion of $O_1^{\mu\nu}$ into the heavy-quark
two-point function vanishes. The remaining divergences can be removed
by adding counterterms proportional to the original operators
$O_1^{\mu\nu}$ and $O_2^{\mu\nu}$ in (\ref{basis}), as well as to the
operators
\begin{eqnarray}
   O_3^{\mu\nu} &=& \bar h_v\Gamma\,(v^\mu Q^\nu - v^\nu Q^\mu)\,
    iv\cdot D\,h_v \nonumber\\[-0.3cm]
   &&\mbox{}+ \bar h_v\,(iv\cdot\overleftarrow{D^\dagger})\,\Gamma\,
    (v^\mu Q^\nu - v^\nu Q^\mu)\,h_v \,, \nonumber\\
   O_4^{\mu\nu} &=& \bar h_v\Gamma \left( [iD^\mu,Q^\nu]
    - [iD^\nu,Q^\mu] \right) h_v \,, \nonumber\\
   O_5^{\mu\nu} &=& v^\mu v_\rho O_4^{\rho\nu}
    - v^\nu v_\rho O_4^{\rho\mu} \,, \nonumber\\
   O_6^{\mu\nu} &=& \bar h_v\Gamma\,[Q^\mu,Q^\nu]\,h_v \,,
    \nonumber\\
   O_7^{\mu\nu} &=& v^\mu v_\rho O_6^{\rho\nu}
    - v^\nu v_\rho O_6^{\rho\mu} \,.
\end{eqnarray}
Here $Q^\mu$ is the `quantum part' of the gluon field, defined by
the decomposition $A^\mu=B^\mu+Q^\mu$. Under a local gauge
transformation, the background and the quantum field transform as
\begin{eqnarray}
   B^\mu(x) &\to& U(x)\,B^\mu(x)\,U^\dagger(x)
    + \frac{i}{g_s}\,[\partial^\mu U(x)]\,U^\dagger(x) \,,
    \nonumber\\
   Q^\mu(x) &\to& U(x)\,Q^\mu(x)\,U^\dagger(x) \,,
\end{eqnarray}
with $U(x)\in SU(N)$, so that the gluon field $A^\mu$ obeys the usual
transformation law. The operators containing the quantum part of the
gluon field are needed as insertions inside loop diagrams. Since the
quantum field $Q^\mu$ transforms in the same way as the covariant
derivative, they are all gauge invariant. The (class-2) operator
$O_3^{\mu\nu}$, which vanishes by the equations of motion, has to be
included since the two-loop calculation is performed off-shell. The
one-loop counterterm coefficients $\delta_n$ are
\begin{equation}
   \delta_1 = (C_A+2 C_F)\,\frac{\alpha_s}{4\pi\epsilon} \,, \qquad
   \delta_2 = 2\delta_3 = \delta_4
   = - C_A\,\frac{\alpha_s}{4\pi\epsilon} \,,
\end{equation}
and $\delta_5=\delta_7=0$. The coefficient $\delta_6$ is irrelevant
for our purposes, since insertions of the operator $O_6^{\mu\nu}$
into one-loop diagrams do not contribute to the Lorentz structure
$v^\alpha v^\nu q^\mu$.

\begin{table}
\caption{\label{tab:counter}
Pole parts proportional to $g_s t_a\Gamma\,v^\alpha v^\nu q^\mu$ of
the one-loop diagrams with counter\-term insertions in units of
$(\alpha_s/4\pi)^2$. The diagram $C_8$ does not contribute.}
\vspace{0.4cm}
\begin{center}
\begin{tabular}{|c|c|c|c|}\hline
\rule[-0.35cm]{0cm}{0.95cm}
 & colour factor
 & $\times\left( \frac{-2\omega}{\mu} \right)^{-2\epsilon}$
 & $\times\left( \frac{-q^2}{\mu^2} \right)^{-\epsilon}$ \\
\hline
$C_1$ & $C_A C_F$ & $-\frac{2}{\epsilon^2}$ & \rule{0cm}{0.5cm}
 \\[0.15cm]
$C_2$ & $C_A^2$ & $-\frac{5}{3\epsilon^2}$ & \\[0.15cm]
      & $C_A T_F\,n_f$ & $\phantom{-}\frac{4}{3\epsilon^2}$ &
 \\[0.15cm]
$C_3$ & $C_A^2$ & $-\frac{5}{3\epsilon^2}$ & \\[0.15cm]
      & $C_A T_F\,n_f$ & $\phantom{-}\frac{4}{3\epsilon^2}$ &
 \\[0.15cm]
$C_4$ & $C_A^2$ & $\phantom{-}\frac{5}{3\epsilon^2}$ & \\[0.15cm]
      & $C_A T_F\,n_f$ & $-\frac{4}{3\epsilon^2}$ & \\[0.15cm]
$C_5$ & $C_A (C_A-2 C_F)$ & $-\frac{1}{\epsilon^2}$ & \\[0.15cm]
$C_6$ & $C_A (C_A-2 C_F)$ & $-\frac{1}{\epsilon^2}$ & \\[0.15cm]
$C_7$ & $C_A C_F$ & $\phantom{-}\frac{2}{\epsilon^2}$ & \\[0.15cm]
$C_9$ & $C_A^2$ & & $\frac{1}{\epsilon^2}
                       + \frac{2}{\epsilon}$ \\[0.15cm]
\hline
\rule{0cm}{0.5cm}
Sum & $C_A^2$ & $-\frac{11}{3\epsilon^2}$
    & $\frac{1}{\epsilon^2} + \frac{2}{\epsilon}$ \\[0.15cm]
    & $C_A C_F$ & $\phantom{-}\frac{4}{\epsilon^2}$ & \\[0.15cm]
    & $C_A T_F\,n_f$ & $\phantom{-}\frac{4}{3\epsilon^2}$ &
\\[0.15cm]
\hline
\end{tabular}
\end{center}
\end{table}

\begin{figure}
\epsfxsize=15cm
\centerline{\epsffile{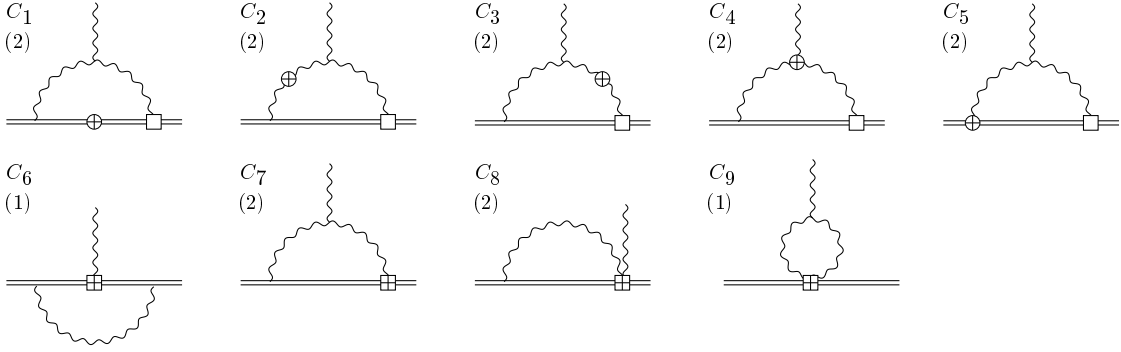}}
\centerline{\parbox{14cm}{\caption{\label{fig:counter}
Counterterm insertion diagrams contributing to the calculation of the
renormalization factor $Z_2$.}}}
\end{figure}

The one-loop diagrams with counterterm insertions, which are required
to cancel the subdivergences of the two-loop diagrams in
Fig.~\ref{fig:2loop}, are shown in Fig.~\ref{fig:counter}. The
resulting pole parts are listed in Tab.~\ref{tab:counter}. Note that
most of the counterterm contributions are pure $1/\epsilon^2$ poles,
since the original diagrams in Fig.~\ref{fig:1loop} do not have
$O(\epsilon^0)$ terms. When the counterterm contributions are added
to the results for the two-loop diagrams, all non-local $1/\epsilon$
divergences proportional to $\ln(-2\omega/\mu)$ and $\ln(-q^2/\mu^2)$
cancel. This is a strong check of our calculation. For the
coefficient of $(\alpha_s/4\pi)^2$, indicated by the subscript in
square brackets, times the tree-level matrix element of
$O_2^{\mu\nu}$, we obtain
\begin{equation}
   - \left( Z_h Z_2 \right)_{[2]} =
   C_A^2 \left( -\frac{4}{3\epsilon^2} + \frac{17}{9\epsilon} \right)
   + C_A C_F\,\frac{2}{\epsilon^2} + C_A T_F\,n_f \left(
   \frac{2}{3\epsilon^2} - \frac{13}{9\epsilon} \right)  \,.
\end{equation}
To obtain $Z_2$, we have to account for the wave-function
renormalization of the heavy-quark fields. (There is no
renormalization of the background field, since $Z_g Z_B^{1/2}$
\cite{Abbo}.) This removes the $C_A C_F$ term. Using then relation
(\ref{nice}) leads to
\begin{equation}
   Z_{\rm mag} = 1 + \frac{C_A\alpha_s}{4\pi\epsilon}
   + \left( \frac{\alpha_s}{4\pi} \right)^2 \Bigg[
   \left( - \frac 43\,C_A^2 + \frac 23\,C_A T_F\,n_f \right)
   \frac{1}{\epsilon^2} +\! \left( \frac{17}{9}\,C_A^2
   - \frac{13}{9}\,C_A T_F\,n_f \right) \frac{1}{\epsilon}\,
   \Bigg] \,.
\end{equation}
The coefficient of the $1/\epsilon^2$ pole in $Z_{\rm mag}$
obeys the relation
\begin{equation}
\label{consistency}
   Z_{{\rm mag},[2]}^{(2)} = \frac 12\,Z_{{\rm mag},[1]}^{(1)}
   \left( Z_{{\rm mag},[1]}^{(1)} - \beta_0 \right) \,,
\end{equation}
where $\beta_0$ is the first coefficient of the $\beta$ function (see
(\ref{coefs}) below), and the super- and subscripts in round (square)
brackets have the same meaning as before. Eq.~(\ref{consistency}) is
a necessary condition for $\gamma_{\rm mag}$ to be finite as
$\epsilon\to 0$ \cite{Flor}. For the anomalous dimension of the
chromo-magnetic operator in the $\overline{\mbox{MS}}$ scheme, we now
obtain from (\ref{gammag})
\begin{equation}
   \gamma_{\rm mag} = \frac{C_A\alpha_s}{2\pi}\left[ 1
   + \left( \frac{17}{18}\,C_A - \frac{13}{18}\,T_F\,n_f \right)
   \frac{\alpha_s}{\pi} + O(\alpha_s^2) \right] \,.
\label{wonder}
\end{equation}
This is our main result.

\section{Conclusions}

Using a relation between renormalization constants that is a
consequence of the re\-parametri\-zation invariance and the virial
theorem, we have found an efficient way to calculate the two-loop
anomalous dimension of the chromo-magnetic operator in the HQET. The
calculation involves the evaluation of 25 non-vanishing two-loop
diagrams. Some of these diagrams have IR divergences that must be
regulated by keeping the external gluon off-shell. After these
divergences are subtracted using a simple variant of the $R^*$
operation, the external momentum can be set to zero and the resulting
two-loop propagator-type tensor integrals can be evaluated using a
recurrence relation, which allows to express them in terms of
products of one-loop integrals.

Our result for the anomalous dimension in (\ref{wonder}) is rather
simple. In our approach, it is obvious from the beginning that the
`abelian' colour structures $C_F^2$ and $C_F T_F\,n_f$ do not appear;
because of the selection rules given in section~\ref{sec:2}, the
result for the anomalous dimension is genuinely non-abelian,
proportional to $C_A$. This would not be obvious if the calculation
were performed in the standard way by studying the mixing of the
chromo-magnetic operator with itself. More surprisingly, there is no
contribution proportional to $C_A C_F$ in the final result, and also
all terms proportional to $\pi^2$ have disappeared. This is in
contrast with the two-loop anomalous dimension of current operators,
where such terms do appear \cite{BrGr,JiMu}--\cite{Kili}.

If, instead of an $SU(N)$ gauge theory, we consider the abelian
$U(1)$ gauge theory, the right-hand side of (\ref{wonder}) vanishes,
since then $C_A=0$ (and $C_F=T_F=1$). In our approach, it is easy to
see that in the abelian theory the anomalous dimension of the
magnetic-moment operator actually vanishes to all orders in
perturbation theory. According to the selection rules, in the absence
of gauge-boson self-couplings the external photon can be connected to
the operator and the heavy-quark lines only through a fermion loop
with at least four photons attached. Such a diagram is convergent
except for subgraphs that correspond to charge and field
renormalization. Consequently, no operator counterterms are needed.

With our two-loop result for $\gamma_{\rm mag}$ at hand, we can
evaluate the Wilson coefficient of the chromo-magnetic operator in
the effective Lagrangian (\ref{Leff}) to next-to-leading order. The
result is
\begin{equation}
   C_{\rm mag}(m_Q/\mu) = \!\left(
\frac{\alpha_s(m_Q)}{\alpha_s(\mu)}
   \right)^{\gamma_0/2\beta_0}\!\left[ 1
   + \frac{\alpha_s(m_Q)}{4\pi}\,c_1 +
   \frac{\alpha_s(m_Q) - \alpha_s(\mu)}{4\pi}\left(
   \frac{\gamma_1}{2\beta_0} - \frac{\gamma_0\beta_1}{2\beta_0^2}
   \right) \!+ \dots \right] \,
\end{equation}
where $c_1 = 2(C_A+C_F)$ is obtained from one-loop matching
\cite{EiHi}. The one- and two-loop coefficients of the anomalous
dimension and $\beta$ function are
\begin{eqnarray}
   \gamma_0 &=& 2 C_A \,, \qquad
    \gamma_1 = \frac{68}{9}\,C_A^2 - \frac{52}{9}\,C_A T_F\,n_f
    \,, \nonumber\\
   \beta_0 &=& \frac{11}{3}\,C_A - \frac 43\,T_F\,n_f \,, \nonumber\\
   \beta_1 &=& \frac{34}{3}\,C_A^2 - \frac{20}{3}\,C_A T_F\,n_f
    - 4 C_F T_F\,n_f \,.
\label{coefs}
\end{eqnarray}
As an application, we discuss the mass splitting between the
ground-state vector and pseudoscalar mesons containing a bottom or a
charm quark. To leading order in the heavy-quark expansion, one finds
\cite{review}
\begin{equation}
   R \equiv \frac{m_{B^*}^2-m_B^2}{m_{D^*}^2-m_D^2}
   = \frac{C_{\rm mag}(m_b/\mu)}{C_{\rm mag}(m_c/\mu)}
   + O(1/m_Q) \,.
\end{equation}
For $N=3$ colours and $n_f=4$ light-quark flavours (as it is
appropriate in the region between the bottom- and charm-quark
masses)
\begin{equation}
   R = \left( \frac{\alpha_s(m_b)}{\alpha_s(m_c)}\right)^{9/25}\,
   \left[ 1 - \frac{7921}{3750}\,
   \frac{\alpha_s(m_c) - \alpha_s(m_b)}{\pi} \right]
   + \Lambda_R \left( \frac{1}{m_c} - \frac{1}{m_b} \right) \,,
\label{Rval}
\end{equation}
where the non-perturbative parameter $\Lambda_R$ accounts for
higher-order corrections in the heavy-quark expansion. Using
$\alpha_s(m_b)=0.22$ and $\alpha_s(m_c)=0.36$, we obtain $R\approx
0.76$, in reasonable agreement with the experimental value $R_{\rm
exp}=0.89\pm 0.01$. The next-to-leading correction in (\ref{Rval})
amounts to minus $10\%$ and deteriorates the comparison with the
experimental value. However, with $\Lambda_R\approx 200\,$MeV
required for agreement with $R_{\rm exp}$, higher-order corrections
in the $1/m_Q$ expansion remain moderate in size.

\vspace{0.3cm}
{\it Acknowledgements:\/}
We are indebted to A.~Grozin for communicating partial results of an
independent calculation of $\gamma_{\rm mag}$ based on a different
technique. One of us (M.N.) would like to thank M.~Ciuchini,
J.~K\"orner and D.~Kreimer for useful discussions. G.A.\ acknowledges
a grant from the Generalitat Valenciana, and partial support by
DGICYT under grant PB94-0080. He also acknowledges the hospitality of
the CERN Theory Division, where this research was performed.

\end{document}